\documentclass[11pt,a4paper]{article}
\usepackage[hyperref]{emnlp2018}
\usepackage{times}
\usepackage{latexsym}

\usepackage{url}

\usepackage{amsmath}
\usepackage{graphicx}
\usepackage{multicol}
\usepackage{multirow}
\usepackage{amssymb}
\usepackage[english]{babel}
\usepackage{colortbl}
\usepackage{amsthm}
\usepackage{footmisc}
\definecolor{mygray}{gray}{.9}
\definecolor{myblack}{gray}{.7}

\usepackage{footmisc}
\setfnsymbol{wiley}

\aclfinalcopy 

\title{Cascaded Mutual Modulation for Visual Reasoning}

\author{Yiqun Yao$^\textrm{{\bf 123}}$, {Jiaming Xu}$^\textrm{{\bf 12}}$\footnote{}~, Feng Wang$^\textrm{{\bf 1}}$  and Bo Xu$^\textrm{{\bf 1234}}$\\
  $^\textrm{1}$Institute of Automation, Chinese Academy of Sciences (CASIA). Beijing, China \\
  $^\textrm{2}$Research Center for Brain-inspired Intelligence, CASIA \\
  $^\textrm{3}$University of Chinese Academy of Sciences \\
  $^\textrm{4}$Center for Excellence in Brain Science and Intelligence Technology, CAS. China \\
  {\tt \{yaoyiqun2014,jiaming.xu,feng.wang,xubo\}@ia.ac.cn} \\}

\date{}

\begin{document}
\maketitle
\let\thefootnote\relax\footnotetext{* Corresponding Author}

\begin{abstract}
Visual reasoning is a special visual question answering problem that is multi-step and compositional by nature, and also requires intensive text-vision interactions. We propose CMM: Cascaded Mutual Modulation as a novel end-to-end visual reasoning model. CMM includes a multi-step comprehension process for both question and image. In each step, we use a Feature-wise Linear Modulation (FiLM) technique to enable textual/visual pipeline to mutually control each other. Experiments show that CMM significantly outperforms most related models, and reach state-of-the-arts on two visual reasoning benchmarks: CLEVR and NLVR, collected from both synthetic and natural languages. Ablation studies confirm that both our multi-step framework and our visual-guided language modulation are critical to the task. Our code is available at \url{https://github.com/FlamingHorizon/CMM-VR}.
\end{abstract}

\section{Introduction}

It is a challenging task in artificial intelligence to perform reasoning with both textual and visual inputs. Visual reasoning task is designed for researches in this field. It is a special visual question answering (VQA) \cite{01_antol2015vqa} problem, requiring a model to infer the relations between entities in both image and text, and generate a textual answer to the question correctly. Unlike other VQA tasks, questions in visual reasoning often contain extensive logical phenomena, and refer to multiple entities, specific attributes and complex relations. Visual reasoning datasets such as CLEVR \cite{02_johnson2017clevr} and NLVR \cite{04_suhr2017corpus} are built on unbiased, synthetic images, with either complex synthetic questions or natural-language descriptions, facilitating in-depth analyses on reasoning ability itself.

\begin{figure}[t]
\begin{center}
\includegraphics[width=6.1cm]{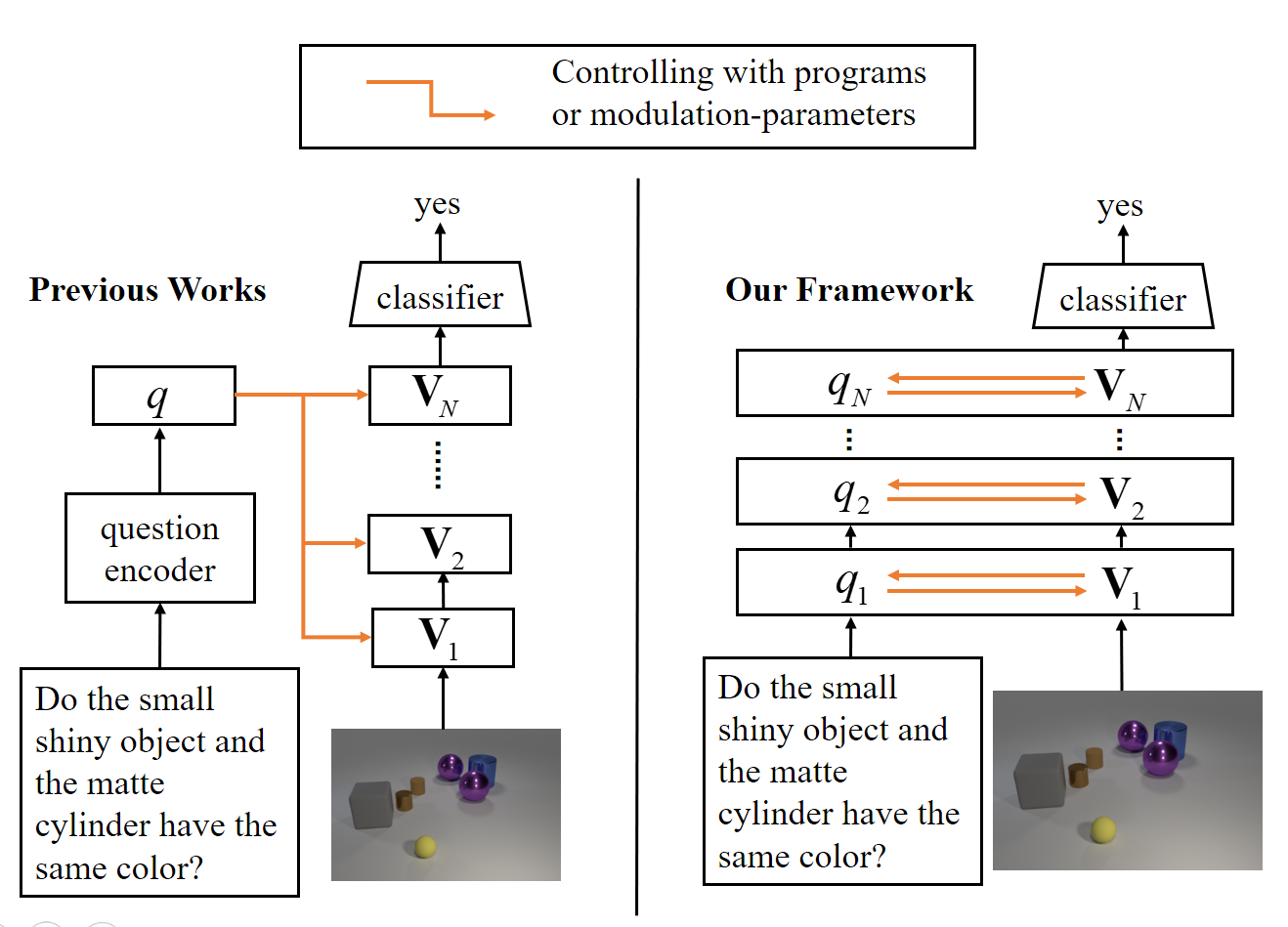}
\caption{Connections and differences between previous ``program-generating'' works and our model: other models generate/control multi-step image-comprehension processes with single question representation, while we put more attention on language logics and let multi-modal information modulate each other in each step. The question and image are taken as a visual-reasoning example from CLEVR dataset.}
\label{figure1}
\end{center}
\end{figure}

Most previous visual reasoning models focus on using the question to guide the multi-step computing on visual features (which can be defined as a image-comprehension ``program''). Neural Module Networks (NMN) \cite{32_andreas2016learning,05_andreas2016neural} and Program Generator + Execution Engine (PG+EE) \cite{03_johnson2017inferring} learn to combine specific image-processing modules, guided by question semantics. Feature-modulating methods like FiLM \cite{07_de2017modulating,08_perez2018film} control image-comprehension process using modulation-parameters generated from the question, allowing models to be trained end-to-end. However, the image-comprehension program in visual reasoning tasks can be extremely long and sophisticated. Using a single question representation to generate or control the whole image-comprehension process raises difficulties in learning. Moreover, since information comes from multiple modalities, it is not intuitive to assume that one (language) is the ``program generator'', and the other (image) is the ``executor''. One way to avoid making this assumption is to perform multiple steps of reasoning with each modality being generator and executor alternately in each step. For these two reasons, we propose Cascaded Mutual Modulation (Figure \ref{figure1}), a novel visual reasoning model to solve the problem that previous ``program-generating'' models lack a method to use visual features to guide multi-step reasoning on language logics. CMM reaches state-of-the-arts on two benchmarks: CLEVR (complex synthetic questions) and NLVR (natural-language).

\section{Related Work}

Perez et al.~\shortcite{08_perez2018film} proposed FiLM as an end-to-end feature-modulating method. The original ResBlock+GRU+FiLM structure uses single question representation, and conditions all image-modulation-parameters on it, without sufficiently handling multi-step language logics. In contrast, we modulate both image and language features alternately in each step, and condition the modulation-parameters on the representations from previous step. We design an image-guided language attention pipeline and use it in combination with FiLM in our CMM framework, and significantly outperform the original structure.

Other widely-cited works on CLEVR/NVLR include Stacked Attention Networks (SAN) \cite{10_yang2016stacked}, NMN \cite{05_andreas2016neural}, N2NMN \cite{09_hu2017learning}, PG+EE \cite{03_johnson2017inferring} and Relation Networks (RN) \cite{23_santoro2017simple}. The recent CAN model \cite{36_hudson2018compositional} also uses multiple question representations and has strong performances on CLEVR. However, these representations are not modulated by the visual part as in our model.

In other VQA tasks, DAN \cite{12_nam2017dual} is the only multi-step dual framework related to ours. For comparison, in every time step, DAN computes textual and visual attention \emph{in parallel} with the same key-vector, while we perform textual attention and visual modulation (instead of attention) in a \emph{cascaded} manner.

\section{Model}

We review and extend FiLM in Section 3.1-3.2, and introduce CMM model in Section 3.3-3.4.

\subsection{Visual Modulation}

Perez et al.~\shortcite{08_perez2018film} proposed \textbf{F}eature-w\textbf{i}se \textbf{L}inear \textbf{M}odulation (FiLM), an affine transformation on intermediate outputs of a neural network ($v$ stands for visual):
\begin{equation}
F{\rm{i}}L{M^v}({{\bf{F}}_{i,c}}|{\gamma _{i,c}},{\beta _{i,c}}) = {\gamma _{i,c}}{{\bf{F}}_{i,c}} + {\beta _{i,c}},
\end{equation}where ${{\bf{F}}_{i,c}}$ is the $c$-th feature map ($C$ in total) generated by Convolutional Neural Networks (CNN) in the $i$-th image-comprehension step. Modulation-parameters ${\gamma _{i,c}}$ and ${\beta _{i,c}}$ can be conditioned on any other part of network (in their work the \emph{single} question representation $q$). If the output tensor ${{\bf{V}}_i}$ of a CNN block is of size $C \times H \times W$, then ${{\bf{F}}_{i,c}}$ is a single slice of size $1 \times H \times W$. $H$ and $W$ are the height and width of each feature map.

Unlike \cite{08_perez2018film}, in each step $i$, we compute a new question vector ${q_i}$. Modulation-parameters ${{\bf{\gamma }}_i}$ and ${{\bf{\beta }}_i}$ ($C \times 1$ vectors, ${{\bf{\gamma }}_i}$ = [$\gamma _{i,1}$$, ... ,$$\gamma _{i,C}$], etc.) are conditioned on the previous question vector ${q_{i-1}}$ instead of a single $q$:
\begin{equation}
{{\bf{\gamma }}_i},{{\bf{\beta }}_i} = ML{P^i}({q_{i - 1}}).
\end{equation}

MLP stands for fully connected layers with linear activations. The weights and biases are not shared among all steps.

\begin{figure*}[ht]
\begin{center}
\includegraphics[width=10.5cm]{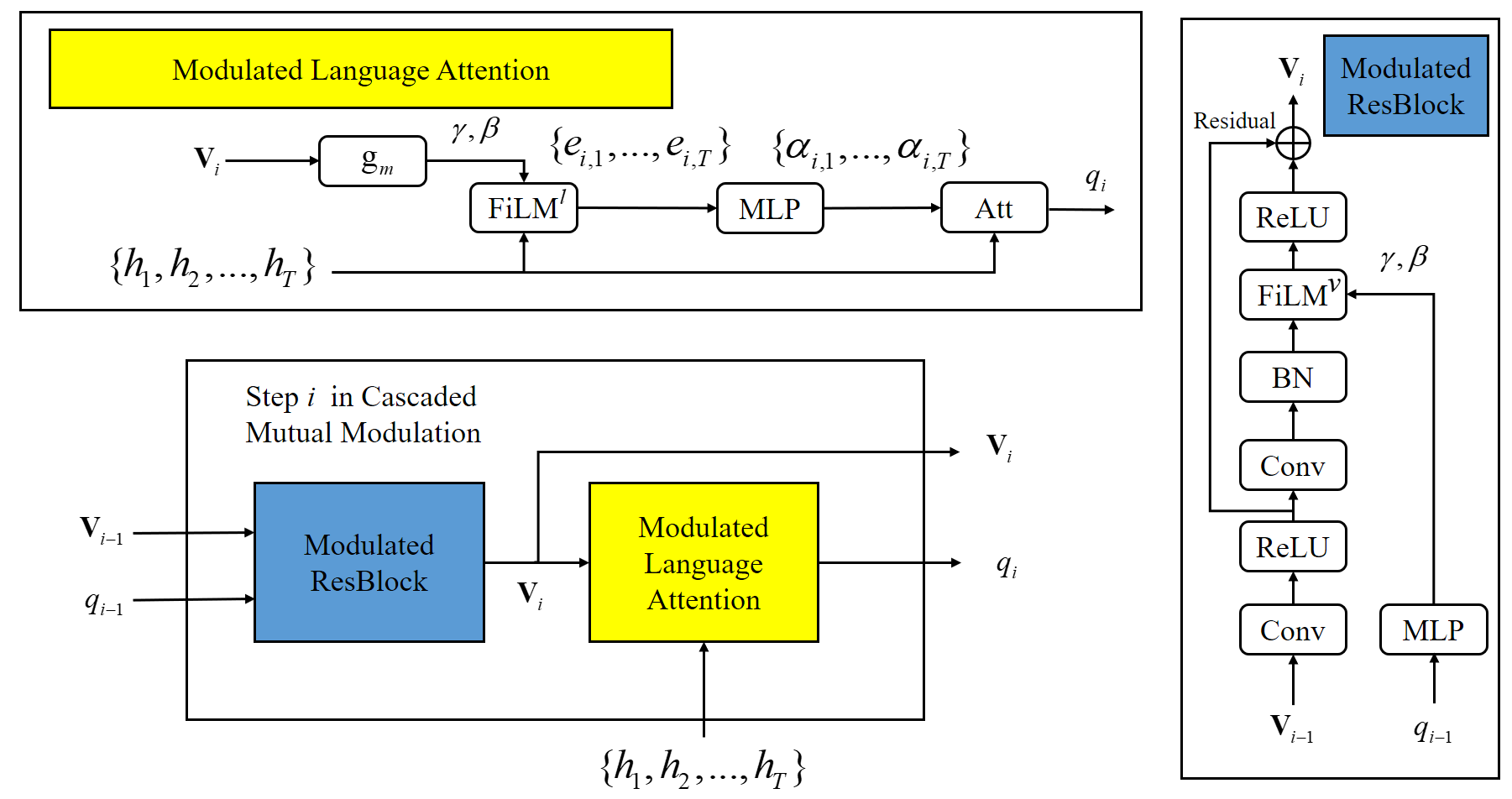}
\caption{Details in CMM step $i$ (middle), with a modulated ResBlock (right) and a modulated textual attention pipeline (top). Visual and textual features modulate each other in each step to compute new representations.}
\label{figure2}
\end{center}
\end{figure*}

\subsection{Language Modulation}

In each step $i$, we also apply FiLM to modulate every language ``feature map''. If the full question representation is a $D \times T$ matrix, a question ``feature map'' ${{\bf{f}}_{i,d}}$ is defined as a $1 \times T$ vector gathering $T$ features along a single dimension. $D$ is the hidden-state dimension of language encoder, and $T$ is a fixed maximum length of word sequences.
\begin{equation}
F{\rm{i}}L{M^l}({{\bf{f}}_{i,d}}|{{\bf{\gamma }}_{i,d}},{{\bf{\beta }}_{i,d}}) = {{\bf{\gamma }}_{i,d}}{{\bf{f}}_{i,d}} + {{\bf{\beta }}_{i,d}},
\end{equation}where $l$ stands for language. Concatenated modulation-parameters ${{\bf{\gamma }}_i}$ and ${{\bf{\beta }}_i}$ ($D \times 1$) are conditioned on the visual features ${{\bf{V}}_i}$ computed in the same step:
\begin{equation}
{{\bf{\gamma }}_i},{{\bf{\beta }}_i} = g_m({{\bf{V}}_i}),
\label{eq4}
\end{equation}where $g_m$ (Section \ref{s3.4}) is an interaction function that converts $3$-d visual features to language-modulation-parameters. The weights in $g_m$ are shared among all $N$ steps.

\subsection{Cascaded Mutual Modulation}

The whole pipeline of our model is built up with multiple \emph{steps}. In each step $i$ ($N$ in total), previous question vector ${q_{i-1}}$ and the visual features ${{\bf{V}}_{i-1}}$ are taken as input; ${q_{i}}$ and ${{\bf{V}}_{i}}$ are computed as output. Preprocessed questions/images are encoded by language/visual encoders to form ${q_{0}}$ and ${{\bf{V}}_{0}}$.

In each step, we cascade a FiLM-ed ResBlock with a modulated textual-attention. We feed ${{\bf{V}}_{i-1}}$ into the ResBlock modulated by parameters from ${q_{i-1}}$ to compute ${{\bf{V}}_{i}}$, and then control the textual attention process with modulation-parameters from ${{\bf{V}}_{i}}$, to compute the new question vector ${q_{i}}$. (Figure \ref{figure2}, middle).

Each ResBlock contains a $1 \times 1$ convolution, a $3 \times 3$ convolution, a batch-normalization \cite{20_ioffe2015batch} layer before FiLM modulation, followed by a residual connection \cite{17_he2016deep}. (Figure \ref{figure2}, right. We keep the same ResBlock structure as \cite{08_perez2018film}). To be consistent with \cite{03_johnson2017inferring,08_perez2018film}, we concatenate the input visual features ${{\bf{V}}_{i-1}}$ of each ResBlock $i$ with two ``coordinate feature maps'' scaled from $-1$ to $1$, to enrich representations of spatial relations. All CNNs in our model use ReLU as activation functions; batch-normalization is applied before ReLU.

After the ResBlock pipeline, we apply language modulation on the full language features $\{ {h_{1}},...,{h_{T}}\}$ ($D \times T$) conditioned on ${{\bf{V}}_{i}}$ and rewrite along the time dimension, yielding:

\begin{equation}
{e_{i,t}} = FiL{M^l}({h_t}|g_m({{\bf{V}}_i})),
\end{equation}
and compute visual-guided attention weights:
\begin{equation}
{\alpha _{i,t}} = {{\mathop{\rm softmax}\nolimits} _t}({\bf{W}}_i^{att}{e_{i,t}} + b_i^{att}),
\label{eq7}
\end{equation}
and weighted summation over time:
\begin{equation}
{q_i} = \sum\limits_{t = 1}^T {{\alpha _{i,t}}{h_t}}.
\end{equation}

In equation (\ref{eq7}), $W_i^{att} \in {\mathbb{R}^{1 \times D}}$ and $b_i^{att} \in {\mathbb{R}^{1 \times 1}}$ are network weights and bias; ${h_t}$ is the $t$-th language state vector ($D \times 1$), computed using a bi-directional GRU \cite{18_chung2014empirical} from word embeddings $\{ {w_1},...,{w_T}\}$. In each step $i$, the language pipeline does not re-compute ${h_t}$, but re-modulate it as ${e_{i,t}}$ instead. (Figure \ref{figure2}, top.)

\begin{table*}[ht]
\small
\centering
\newcommand{\tabincell}[2]{\begin{tabular}{@{}#1@{}}#2\end{tabular}}
\begin{tabular}{lcccccc}
\hline
Model & Overall & Count & Exist & \tabincell{c}{Compare \\ Numbers} & \tabincell{c}{Query \\ Attribute} & \tabincell{c}{Compare \\ Attribute} \\
\hline
Human & 92.6 & 86.7 & 96.6 & 86.5 & 95.0 & 96.0\\\hline
SAN \cite{10_yang2016stacked} & 76.7 & 64.4 & 82.7 & 77.4 & 82.6 & 75.4 \\
N2NMN \cite{32_andreas2016learning} & 83.7 & 68.5 & 85.7 & 84.9 & 90.0 & 88.7\\
PG+EE-9K & 88.6 & 79.7 & 89.7 & 79.1 & 92.6 & 96.0\\
PG+EE-700K \cite{03_johnson2017inferring} & 96.9 & 92.7 & 97.1 & 98.7 & 98.1 & 98.9\\
RN \cite{23_santoro2017simple} & 95.5 & 90.1 & 97.8 & 93.6 & 97.9 & 97.1\\
COG-model \cite{38_yang2018dataset} & 96.8 & 91.7 & 99.0 & 95.5 & 98.5 & 98.8 \\
FiLM & 97.7 & 94.3 & 99.1 & 96.8 & 99.1 & 99.1\\
FiLM-raw \cite{08_perez2018film} & 97.6 & 94.3 & 99.3 & 93.4 & 99.3 & 99.3\\
DDRprog \cite{35_suarez2018ddrprog} & 98.3 & 96.5 & 98.8 & 98.4 & 99.1 & 99.0 \\
CAN \cite{36_hudson2018compositional} & 98.9 & 97.1 & \bf{99.5} & \bf{99.1} & 99.5 & \bf{99.5} \\
\hline
CMM-single (ours) & 98.6 & 96.8 & 99.2 & 97.7 & 99.4 & 99.1\\
CMM-ensemble (ours) & \bf{99.0} & \bf{97.6} & \bf{99.5} & 98.5 & \bf{99.6} & 99.4\\
\hline
\end{tabular}

\caption{Accuracies on CLEVR test set. N2NMN and PG+EE need extra supervision to train with reinforcement learning. FiLM-raw uses raw image as input (others use pre-extracted features). Another work \cite{37_mascharka2018transparency} gets 99.1\% accuracy but uses strong program supervision, which is a totally different setting.}

\label{table1}

\end{table*}

\subsection{Feature Projections}
\label{s3.4}

We use a function $g_p$ to project the last visual features ${{\bf{V}}_N}$ into a final representation:
\begin{equation}
{u_{final}} = g_p({{\bf{V}}_N}).
\end{equation}

$g_p$ includes a convolution with $K$ $1 \times 1$ kernels, a batch-normalization afterwards, followed by global max pooling over all pixels ($K$ = $512$).

We also need a module $g_m$ (equation (\ref{eq4})) to compute language-modulations with ${{\bf{V}}_{i}}$, since ${{\bf{V}}_{i}}$ is 3-d features (not a weighted-summed vector as in traditional visual-attention). We choose $g_m$ to have the same structure as $g_p$, except that $K$ equals to the total number of modulation-parameters in each step. This design is critical (Section \ref{s4.3}).

We use a fully connected layer with $1024$ ReLU hidden units as our answer generator. It takes $u_{final}$ as input, and predicts the most probable answer in the answer vocabulary.

\section{Experiments}

We are the first to achieve top results on both datasets (CLEVR, NLVR) with one structure. See Appendix for more ablation and visualization results.

\subsection{CLEVR}

CLEVR \cite{02_johnson2017clevr} is a commonly-used visual reasoning benchmark containing 700,000 training samples, 150,000 for validation and test. Questions in CLEVR cover several typical elements of reasoning: counting, comparing, querying the memory, etc. Many well-designed models on VQA have failed on CLEVR, revealing the difficulty to handle the multi-step and compositional nature of logical questions.

On CLEVR dataset, we embed the question words into a 200-dim continuous space, and use a bi-directional GRU with 512 hidden units to generate 1024-dim question representations. Questions are padded with NULL token to a maximum length $T = 46$. As the first-step question vector in CMM, ${q_0}$ can be arbitrary RNN hidden state in the set $\{ {h_1},...,{h_T}\} $ (Section 3.3). We choose the one at the end of the unpadded question.

In each ResBlock, the feature map number $C$ is set to 128. Images are pre-processed with a ResNet101 network pre-trained on ImageNet \cite{22_russakovsky2015imagenet} to extract $1024 \times 14 \times 14$ visual features (this is also common practice on CLEVR). We use a trainable one-layer CNN with 128 kernels ($3 \times 3$) to encode the extracted features into ${{\bf{V}}_{0}}$ ($128 \times 14 \times 14$). Convolutional paddings are used to keep the feature map size to be $14 \times 14$ through the visual pipeline.

We train the model with an ADAM \cite{19_kingma2014adam} optimizer using a learning rate of 2.5e-4 and a batch-size of 64 for about 90 epoches, and switch to an SGD with the same learning rate and 0.9 momentum, fine-tuning for another 20 epoches. SGD generally brings around 0.3 points gains to CMM on CLEVR.

We achieve 98.6\% accuracy with single model (4-step), significantly better than FiLM and other related work, only slightly lower than CAN, but CAN needs at least 8 model-blocks for $>$98\% (and 12 for best). We achieve state-of-the-art of 99.0\% with ensemble of 4/5/6 step CMM models. Table \ref{table1} shows test accuracies on all types of questions. The main improvements over program-generating models come from ``Counting'' and ``Compare Numbers'', indicating that CMM framework significantly enhances language (especially numeric) reasoning without sophisticated memory design like CAN.

\subsection{NLVR}
NLVR \cite{04_suhr2017corpus} is a visual reasoning dataset proposed by researchers in NLP field. NLVR has 74,460 samples for training, 5,940 for validation and 5,934 for public test. In each sample, there is a human-posed natural language description on an image with 3 sub-images, and requires a false/true response.

We use different preprocessing methods on NLVR. Before training, we reshape NLVR images into $14 \times 56$ raw pixels and use them directly as visual inputs ${{\bf{V}}_{0}}$. For language part, we correct some obvious typos among the rare words (frequency $<$ 5) in the training set, and pad the sentences to a maximum length of 26. Different from CLEVR, LSTM works better than GRU on the real-world questions. For training, we use ADAM with a learning rate of 3.5e-4 and a batch-size of 128 for about 200 epoches, without SGD fine-tuning.

Our model (3-step, 69.9\%) outperforms all proposed models on both validation and public test set\footnote{According to NLVR rules, we will run on the unreleased test set (Test-U) in the near future.}, showing that CMM is also suitable for real-world languages (Table \ref{table3}).

\begin{table}[ht]
\centering
\small
\newcommand{\tabincell}[2]{\begin{tabular}{@{}#1@{}}#2\end{tabular}}
\begin{tabular}{lccc}
\hline
Model  & \tabincell{c}{Dev} & \tabincell{c}{Test-P}\\\hline
Text only  & 56.6 & 57.2\\
Image only  & 55.4 & 56.1\\
CNN+RNN \cite{04_suhr2017corpus}  & 56.6 & 58.0\\
NMN \cite{05_andreas2016neural} & 63.1 & 66.1\\
FiLM (our run)  & 59.0 & 61.3\\
CNN-BiAtt \cite{39_tan2018object} & 66.9 & 69.7\\
\hline
CMM-3-steps (ours) & \bf{68.0} & \bf{69.9}\\\hline
\end{tabular}

\caption{Accuracies on valid and test set of NLVR.}
\label{table3}

\end{table}

\subsection{Ablation Studies}

\label{s4.3}
We list CMM ablation results in Table \ref{table4}. Ablations on CLEVR show that CMM is robust to step number but sensitive to $g_m$ structure because it's the key to multi-modal interaction. Section \ref{s3.4} is temporarily a best choice. CMM performs over 7-point higher than FiLM on NLVR in a setting of same hyper-parameters and ResBlocks, showing the importance of handling \emph{language} logics (see also difficult CLEVR subtasks in Table \ref{table1}).

\begin{table}[h]
\centering
\small
\newcommand{\tabincell}[2]{\begin{tabular}{@{}#1@{}}#2\end{tabular}}
\begin{tabular}{cc|ccc}
\hline
\multicolumn{2}{c|}{CLEVR} & \multicolumn{3}{c}{NLVR}\\\hline
Model & Val & Model & Dev & \tabincell{c}{Test-P}\\\hline
5-step & 98.4 & FiLM-hyp & 59.0 & 61.3\\
6-step & 98.4 & 1-step & 65.3 & 66.9\\
$g_m$-CNN & 93.3 & 2-step & 67.7 & 66.8\\
$g_m$ w/o BN & 94.1 & 3-step & 68.5 & 68.4\\
$g_m$-NS & 97.0 & 4-step & 67.2 & 66.5\\
\hline
4-step & 98.6 & 3-step-LSTM & 68.0 & 69.9\\\hline
\end{tabular}

\caption{Ablation studies on CLEVR/NLVR. $g_m$-CNN means using 2-layer-CNN with 3 $\times$ 3 kernels, followed by concatenation and MLP, as $g_m$. BN means batch-normalization in $g_m$. NS means not sharing weights. ``FiLM-hyp'' uses all the same hyper-parameters as the 3-step CMM (both use GRU as question encoder).}
\label{table4}

\end{table}

\begin{table}[ht]
\centering
\tiny
\begin{tabular}{|cccc|c|}
\hline
Words & Block 1 & Block 2 & Block 3 & Visual Attention Map\\
\hline
$<$START$>$ & 0.017 & 0.030 & \multicolumn{1}{>{\columncolor{mygray}}c}{0.052} & \multirow{8}{*}{\includegraphics[height=1.9cm ,width=0.3\columnwidth]{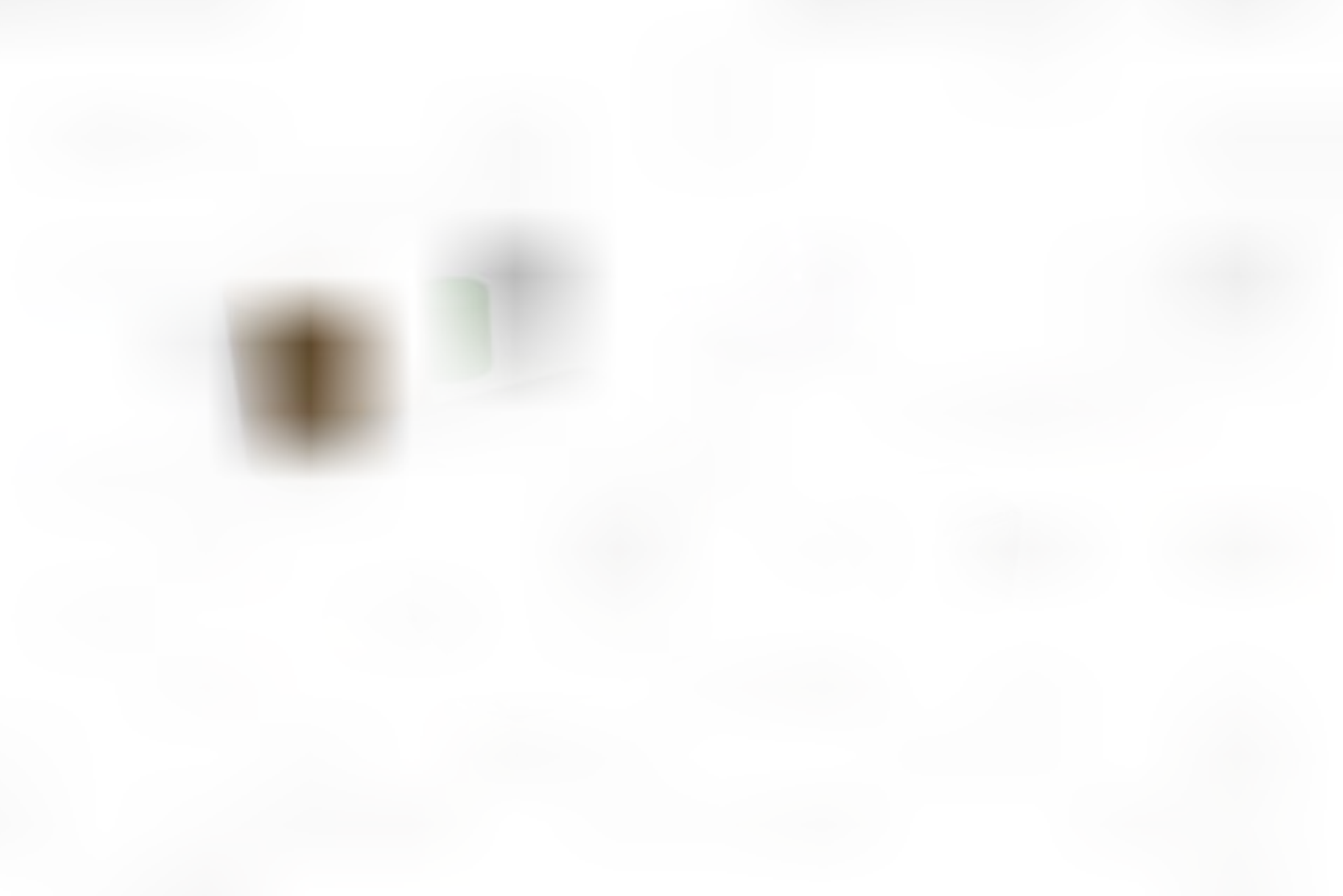}}\\
Is & 0.006 & 0.048 & \multicolumn{1}{>{\columncolor{myblack}}c}{0.501} & \\
there & 0.010 & \multicolumn{1}{>{\columncolor{mygray}}c}{0.041} & \multicolumn{1}{>{\columncolor{mygray}}c}{0.299} &\\
a & 0.006 & 0.031 & \multicolumn{1}{>{\columncolor{mygray}}c}{0.015} & \\
big & 0.008 & 0.031 & 0.001 & \\
brown & \multicolumn{1}{>{\columncolor{mygray}}c}{0.049} & 0.040 & 0.002 & \\
object & \multicolumn{1}{>{\columncolor{mygray}}c}{0.149} & \multicolumn{1}{>{\columncolor{myblack}}c}{0.265} & 0.003 & \\
of & 0.032 & \multicolumn{1}{>{\columncolor{mygray}}c}{0.063} & 0.003 & \\
the & 0.030 & 0.029 & 0.004 & \multirow{8}{*}{\includegraphics[height=1.9cm ,width=0.3\columnwidth]{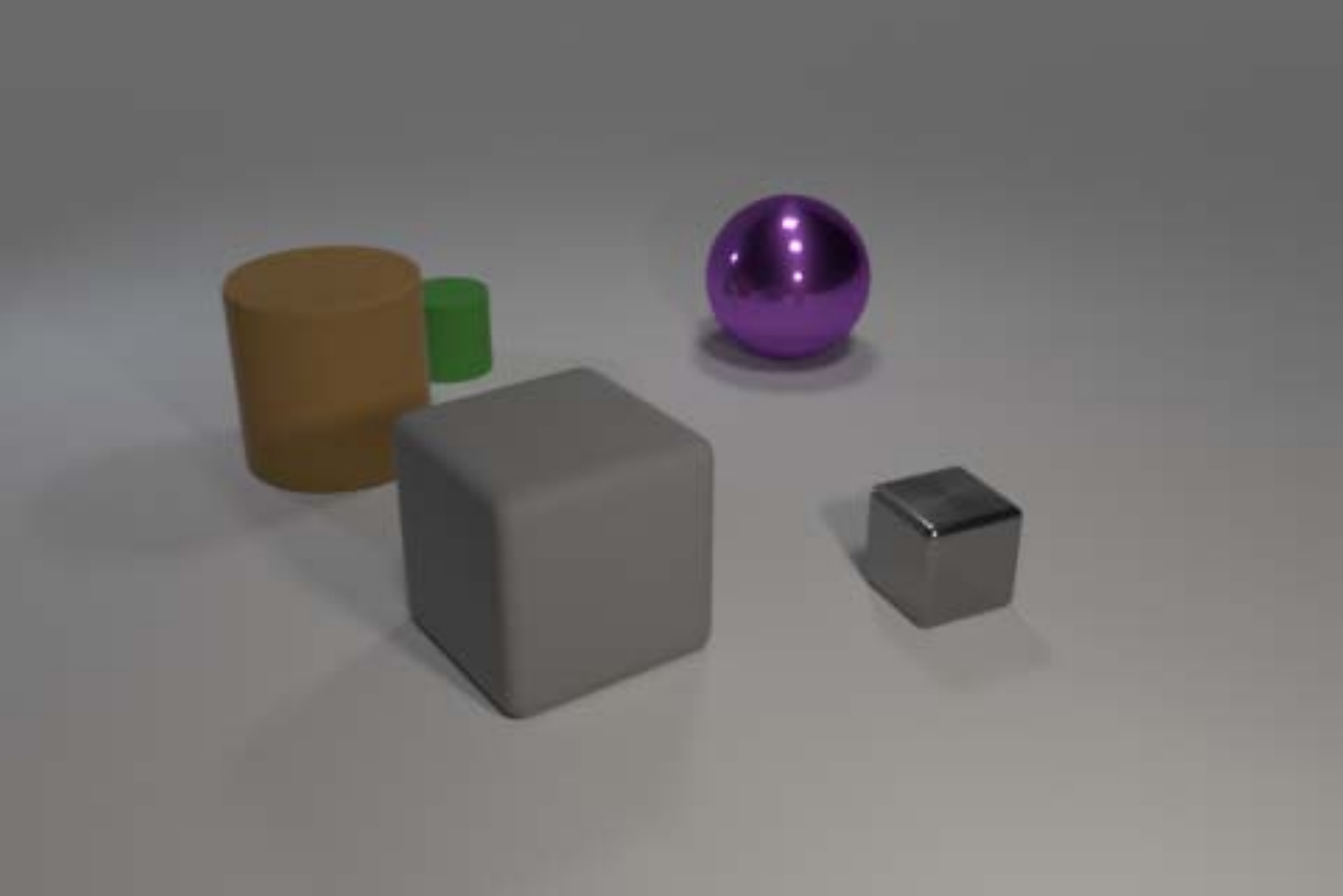}}\\
same & \multicolumn{1}{>{\columncolor{mygray}}c}{0.062} & 0.032 & 0.002 & \\
shape & \multicolumn{1}{>{\columncolor{myblack}}c}{0.345} & \multicolumn{1}{>{\columncolor{mygray}}c}{0.152} & 0.007 & \\
as & \multicolumn{1}{>{\columncolor{mygray}}c}{0.233} & 0.031 & 0.001 & \\
the & 0.000 & 0.005 & 0.004 & \\
green & 0.003 & 0.014 & 0.004 & \\
thing & 0.012 & \multicolumn{1}{>{\columncolor{mygray}}c}{0.075} & 0.011 & \\
$<$END$>$ & 0.006 & 0.036 & \multicolumn{1}{>{\columncolor{mygray}}c}{0.025} & \\\hline

\end{tabular}

\caption{Visualization of CMM intermediate outputs on a sample from CLEVR validation set. We colour the largest attention weight with dark gray, and top four attention weights in the rest with light gray.}
\label{table5}
\end{table}

\subsection{A Case Study}

We select an image-question pair from the validation set of CLEVR for visualization. In Table \ref{table5}, we visualize the multi-step attention weights on question words, and the distribution of argmax position in the global max-pooling layer of $g_p$ (equivalent to the last visual ``attention map'' although there isn't explicit visual attention in our image-comprehension pipeline). On the bottom right is the original image, and on the top right is the distribution of argmax positions in the global max-pooling, multiplied by the original image.

Our model attends to phrases ``same shape as'' and ``brown object'' in the first two reasoning steps. These phrases are meaningful because ``same shape as'' is the core logic in the question, and ``brown object'' is the key entity to generating the correct answer. In the last step, the model attends to the phrase ``is there''. This implicitly classifies the question into question-type ``exist'', and directs the answer generator to answer ``no'' or ``yes''. The visual map, guided by question-based modulation parameters, concentrates on the green and brown object correctly.

The result shows that visual features can guide the comprehension of question logics with textual modulation. On the other hand, question-based modulation parameters enable the ResBlocks to filter out irrelative objects.

\section{Conclusion}

We propose CMM as a novel visual reasoning model cascading visual and textual modulation in each step. CMM reaches state-of-the-arts on visual reasoning benchmarks with both synthetic and real-world languages.

\section*{Acknowledgements}
We thank the reviewers for their insightful comments. This work is supported by the National Natural Science Foundation of China (61602479) and the Strategic Priority Research Program of the Chinese Academy of Sciences (XDBS01070000).

\bibliography{submit_01}

\begin{thebibliography}{22}
\expandafter\ifx\csname natexlab\endcsname\relax\def\natexlab#1{#1}\fi

\bibitem[{Andreas et~al.(2016{\natexlab{a}})Andreas, Rohrbach, Darrell, and
  Klein}]{32_andreas2016learning}
Jacob Andreas, Marcus Rohrbach, Trevor Darrell, and Dan Klein.
  2016{\natexlab{a}}.
\newblock Learning to compose neural networks for question answering.
\newblock In \emph{Proceedings of NAACL-HLT}, pages 1545--1554.

\bibitem[{Andreas et~al.(2016{\natexlab{b}})Andreas, Rohrbach, Darrell, and
  Klein}]{05_andreas2016neural}
Jacob Andreas, Marcus Rohrbach, Trevor Darrell, and Dan Klein.
  2016{\natexlab{b}}.
\newblock Neural module networks.
\newblock In \emph{Proceedings of the IEEE Conference on Computer Vision and
  Pattern Recognition}, pages 39--48.

\bibitem[{Antol et~al.(2015)Antol, Agrawal, Lu, Mitchell, Batra,
  Lawrence~Zitnick, and Parikh}]{01_antol2015vqa}
Stanislaw Antol, Aishwarya Agrawal, Jiasen Lu, Margaret Mitchell, Dhruv Batra,
  C~Lawrence~Zitnick, and Devi Parikh. 2015.
\newblock Vqa: Visual question answering.
\newblock In \emph{Proceedings of the IEEE International Conference on Computer
  Vision}, pages 2425--2433.

\bibitem[{Chung et~al.(2014)Chung, Gulcehre, Cho, and
  Bengio}]{18_chung2014empirical}
Junyoung Chung, Caglar Gulcehre, KyungHyun Cho, and Yoshua Bengio. 2014.
\newblock Empirical evaluation of gated recurrent neural networks on sequence
  modeling.
\newblock \emph{arXiv preprint arXiv:1412.3555}.

\bibitem[{De~Vries et~al.(2017)De~Vries, Strub, Mary, Larochelle, Pietquin, and
  Courville}]{07_de2017modulating}
Harm De~Vries, Florian Strub, J{\'e}r{\'e}mie Mary, Hugo Larochelle, Olivier
  Pietquin, and Aaron~C Courville. 2017.
\newblock Modulating early visual processing by language.
\newblock In \emph{Advances in Neural Information Processing Systems}, pages
  6597--6607.

\bibitem[{He et~al.(2016)He, Zhang, Ren, and Sun}]{17_he2016deep}
Kaiming He, Xiangyu Zhang, Shaoqing Ren, and Jian Sun. 2016.
\newblock Deep residual learning for image recognition.
\newblock In \emph{Proceedings of the IEEE conference on computer vision and
  pattern recognition}, pages 770--778.

\bibitem[{Hu et~al.(2017)Hu, Andreas, Rohrbach, Darrell, and
  Saenko}]{09_hu2017learning}
Ronghang Hu, Jacob Andreas, Marcus Rohrbach, Trevor Darrell, and Kate Saenko.
  2017.
\newblock Learning to reason: End-to-end module networks for visual question
  answering.
\newblock \emph{CoRR, abs/1704.05526}, 3.

\bibitem[{Hudson and Manning(2018)}]{36_hudson2018compositional}
Drew~A Hudson and Christopher~D Manning. 2018.
\newblock Compositional attention networks for machine reasoning.
\newblock In \emph{International Conference on Learning Representations}.

\bibitem[{Ioffe and Szegedy(2015)}]{20_ioffe2015batch}
Sergey Ioffe and Christian Szegedy. 2015.
\newblock Batch normalization: Accelerating deep network training by reducing
  internal covariate shift.
\newblock In \emph{International conference on machine learning}, pages
  448--456.

\bibitem[{Johnson et~al.(2017{\natexlab{a}})Johnson, Hariharan, van~der Maaten,
  Fei-Fei, Zitnick, and Girshick}]{02_johnson2017clevr}
Justin Johnson, Bharath Hariharan, Laurens van~der Maaten, Li~Fei-Fei,
  C~Lawrence Zitnick, and Ross Girshick. 2017{\natexlab{a}}.
\newblock Clevr: A diagnostic dataset for compositional language and elementary
  visual reasoning.
\newblock In \emph{IEEE Conference on Computer Vision and Pattern Recognition
  (CVPR), 2017}, pages 1988--1997. IEEE.

\bibitem[{Johnson et~al.(2017{\natexlab{b}})Johnson, Hariharan, van~der Maaten,
  Hoffman, Fei-Fei, Zitnick, and Girshick}]{03_johnson2017inferring}
Justin Johnson, Bharath Hariharan, Laurens van~der Maaten, Judy Hoffman,
  Li~Fei-Fei, C~Lawrence Zitnick, and Ross Girshick. 2017{\natexlab{b}}.
\newblock Inferring and executing programs for visual reasoning.
\newblock In \emph{Proceedings of the IEEE International Conference on Computer
  Vision}.

\bibitem[{Kingma and Ba(2014)}]{19_kingma2014adam}
Diederik~P Kingma and Jimmy Ba. 2014.
\newblock Adam: A method for stochastic optimization.
\newblock \emph{arXiv preprint arXiv:1412.6980}.

\bibitem[{Mascharka et~al.(2018)Mascharka, Tran, Soklaski, and
  Majumdar}]{37_mascharka2018transparency}
David Mascharka, Philip Tran, Ryan Soklaski, and Arjun Majumdar. 2018.
\newblock Transparency by design: Closing the gap between performance and
  interpretability in visual reasoning.
\newblock In \emph{Proceedings of the IEEE conference on computer vision and
  pattern recognition}.

\bibitem[{Nam et~al.(2017)Nam, Ha, and Kim}]{12_nam2017dual}
Hyeonseob Nam, Jung-Woo Ha, and Jeonghee Kim. 2017.
\newblock Dual attention networks for multimodal reasoning and matching.
\newblock In \emph{Proceedings of the IEEE Conference on Computer Vision and
  Pattern Recognition}, pages 299--307.

\bibitem[{Perez et~al.(2018)Perez, Strub, De~Vries, Dumoulin, and
  Courville}]{08_perez2018film}
Ethan Perez, Florian Strub, Harm De~Vries, Vincent Dumoulin, and Aaron
  Courville. 2018.
\newblock Film: Visual reasoning with a general conditioning layer.
\newblock In \emph{Proceedings of the 32nd AAAI Conference on Artificial
  Intelligence}.

\bibitem[{Russakovsky et~al.(2015)Russakovsky, Deng, Su, Krause, Satheesh, Ma,
  Huang, Karpathy, Khosla, Bernstein et~al.}]{22_russakovsky2015imagenet}
Olga Russakovsky, Jia Deng, Hao Su, Jonathan Krause, Sanjeev Satheesh, Sean Ma,
  Zhiheng Huang, Andrej Karpathy, Aditya Khosla, Michael Bernstein, et~al.
  2015.
\newblock Imagenet large scale visual recognition challenge.
\newblock \emph{International Journal of Computer Vision}, 115(3):211--252.

\bibitem[{Santoro et~al.(2017)Santoro, Raposo, Barrett, Malinowski, Pascanu,
  Battaglia, and Lillicrap}]{23_santoro2017simple}
Adam Santoro, David Raposo, David~G Barrett, Mateusz Malinowski, Razvan
  Pascanu, Peter Battaglia, and Tim Lillicrap. 2017.
\newblock A simple neural network module for relational reasoning.
\newblock In \emph{Advances in neural information processing systems}, pages
  4974--4983.

\bibitem[{Suarez et~al.(2018)Suarez, Johnson, and Li}]{35_suarez2018ddrprog}
Joseph Suarez, Justin Johnson, and Fei-Fei Li. 2018.
\newblock Ddrprog: A clevr differentiable dynamic reasoning programmer.
\newblock \emph{arXiv preprint arXiv:1803.11361}.

\bibitem[{Suhr et~al.(2017)Suhr, Lewis, Yeh, and Artzi}]{04_suhr2017corpus}
Alane Suhr, Mike Lewis, James Yeh, and Yoav Artzi. 2017.
\newblock A corpus of natural language for visual reasoning.
\newblock In \emph{Proceedings of the 55th Annual Meeting of the Association
  for Computational Linguistics (Volume 2: Short Papers)}, volume~2, pages
  217--223.

\bibitem[{Tan and Bansal(2018)}]{39_tan2018object}
Hao Tan and Mohit Bansal. 2018.
\newblock Object ordering with bidirectional matchings for visual reasoning.
\newblock In \emph{Proceedings of NAACL-HLT}.

\bibitem[{Yang et~al.(2018)Yang, Ganichev, Wang, Shlens, and
  Sussillo}]{38_yang2018dataset}
Guangyu~Robert Yang, Igor Ganichev, Xiao-Jing Wang, Jonathon Shlens, and David
  Sussillo. 2018.
\newblock A dataset and architecture for visual reasoning with a working
  memory.
\newblock \emph{arXiv preprint arXiv:1803.06092}.

\bibitem[{Yang et~al.(2016)Yang, He, Gao, Deng, and Smola}]{10_yang2016stacked}
Zichao Yang, Xiaodong He, Jianfeng Gao, Li~Deng, and Alex Smola. 2016.
\newblock Stacked attention networks for image question answering.
\newblock In \emph{Proceedings of the IEEE Conference on Computer Vision and
  Pattern Recognition}, pages 21--29.

\end{thebibliography}
\bibliographystyle{acl_natbib_nourl}

\end{document}